\documentclass[notitlepage,11pt,showkeys]{revtex4-1}

\usepackage{multibib}

\usepackage{cmbright,lmodern}

\usepackage[english]{babel}

\usepackage{amssymb}
\usepackage{xspace}
\usepackage{amsmath}
\usepackage{graphicx}
\usepackage{floatpag}

\newcommand{\bs}{\boldsymbol}
\newcommand{\re}{\operatorname{Re}}
\newcommand{\im}{\operatorname{Im}}
\newcommand{\rmi}{\operatorname{i}}

\newcommand{\mr}{\mathrm}

\newcommand{\rmd}{\mathrm{d}}

\newcommand{\HG}{\mathrm{HG}}
\newcommand{\LG}{\mathrm{LG}}
\newcommand{\GG}{\mathrm{GG}}

\begin{document}

\title{Swings and roundabouts: \\ Optical Poincar\'e spheres for polarization and Gaussian beams}

\author{Mark R Dennis}

\affiliation{H H Wills Physics Laboratory, University of Bristol, Bristol BS8 1TL, UK}

\author{Miguel A Alonso}

\affiliation{The Institute of Optics, University of Rochester, Rochester, New York 14627, USA}

\begin{abstract}
The connection between Poincar\'e spheres for polarization and Gaussian beams is explored, focusing on the interpretation of elliptic polarization in terms of the isotropic 2-dimensional harmonic oscillator in Hamiltonian mechanics, its canonical quantization and semiclassical interpretation.
This leads to the interpretation of structured Gaussian modes, the Hermite-Gaussian, Laguerre-Gaussian and Generalized Hermite-Laguerre Gaussian modes as eigenfunctions of operators corresponding to the classical constants of motion of the 2-dimensional oscillator, which acquire an extra significance as families of classical ellipses upon semiclassical quantization. 
\end{abstract}


\maketitle

\section{Introduction}

There are many analogies between quantum mechanics and classical optical physics, not least because both involve the mathematical theory of waves including ideas such as Fourier transforms, cavity modes, etc.
Since these concepts may appear more natural and fundamental in a quantum setting, wave optical phenomena are often presented in quantum mechanical terms, sometimes as special cases of their quantum mechanical counterparts, such as the interpretation of the bandwidth theorem by Heisenberg's uncertainty relations \cite{Gabor:1947theory}, functions in Hilbert spaces representing quantum states and optical fields, and the free-space paraxial equation as Schr\"odinger's equation, with propagation distance as time.
Even without the notion of photons and field quantization, wave mechanics underpin both wave optics and quantum mechanics.

Here, we take the opportunity to explore another such analogy: the connection between polarization---states of elliptic polarization, as parametrized by Stokes parameters and the Poincar\'e sphere---and the analogous representations of high-order Gaussian laser modes, especially the celebrated Hermite-Gauss (HG) and Laguerre-Gauss (LG) mode sets.

\begin{figure}
\centering\includegraphics[width=\textwidth]{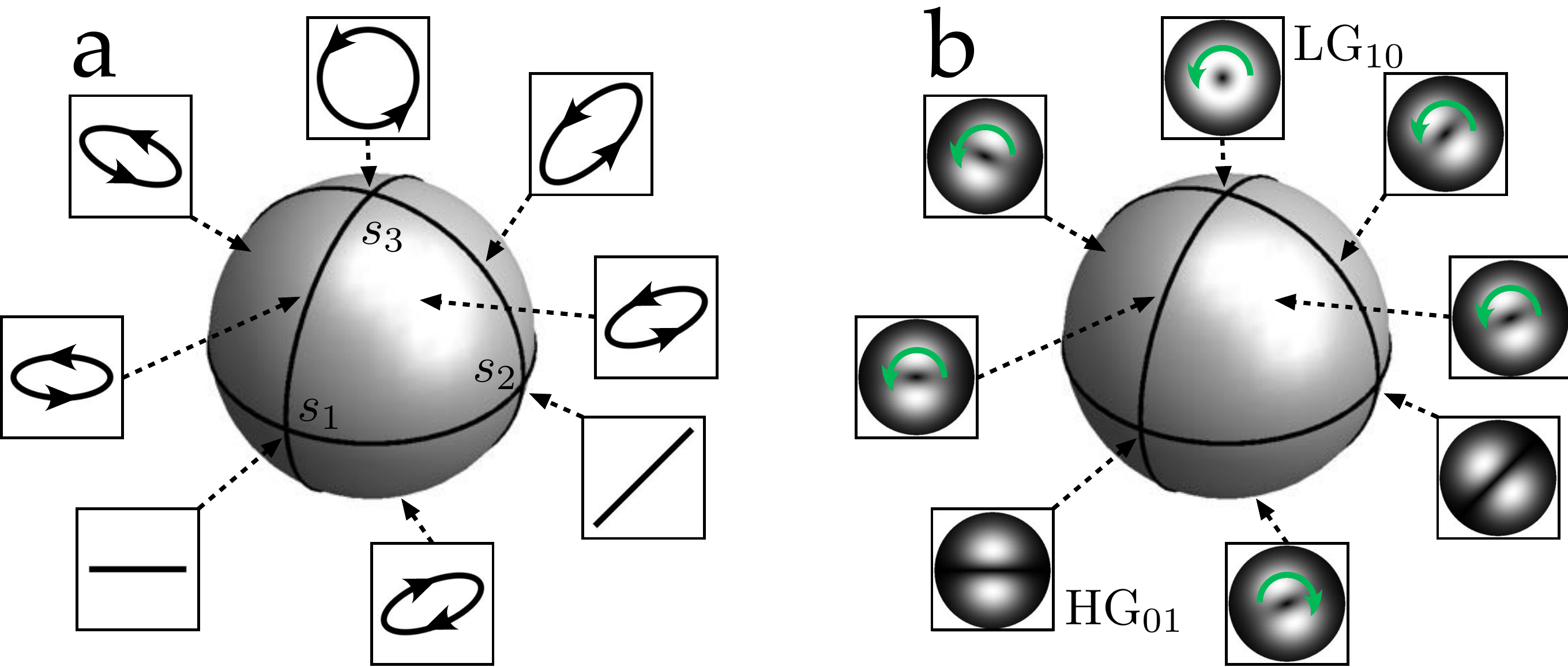}
    \caption{Illustration of the Poincar\'e spheres for elliptic polarization and of Gaussian beams of mode order 1.
    (a) Polarization ellipses.  
    The ellipses have a right-handed sense in the northern hemisphere, left-handed in the southern, with circular polarizations at the poles and linear polarization on the equator.
    The axis angle of the ellipse is half the azimuth angle on the sphere.
    (b) Gaussian beams. 
    Rotated HG modes occur on the equator instead of linear polarizations, and LG vortex modes (of positive or negative sign) occur at the poles instead of linear polarization.
    The analogues of elliptical polarization have a single vortex on axis with an elliptical core.
     }
    \label{fig:poincare}
\end{figure}

The similarity between the Poincar\'e sphere parametrization of linear, circular and elliptic states of polarization, and HG, LG as well as the less familiar `Generalized Hermite-Laguerre-Gaussian' (GG) modes \cite{Danakas:1992analogies,Abramochkin:2004generalized,Visser:2004oam}, has been much explored over the last 25 years, following the important observation of the `equivalence' of the Poincar\'e sphere for polarization and Gaussian modes of mode order unity by Padgett and Courtial \cite{Padgett:1999poincare}, as shown in Figure \ref{fig:poincare}.
We will describe how this analogy emerges naturally, from interpreting polarization in terms of the Hamiltonian mechanics of an isotropic 2-dimensional (2D) oscillator, and Gaussian modes from its canonical quantization in terms of operators \cite{Stoler:1981operator,VanEnk:1992eigenfunction}.
The notion of angular momentum, both in the sense of the spin angular momentum encapsulated by the third Stokes parameter $S_3$, and the orbital angular momentum operator of which the LG modes are eigenfunctions, plays a central role in this picture.

Many aspects of the story we tell have been described before in some detail, especially in \cite{Danakas:1992analogies,Nienhuis:1993paraxial,Abramochkin:2004generalized,Visser:2004oam,Calvo:2005wigner,habraken:thesis}, although we draw stronger mathematical analogies between the classical Hamiltonian structure of the Poincar\'e sphere for polarization, its formal canonical quantization for Gaussian beams, and the semiclassical relationship between the two.
The `swings and roundabouts' nature of harmonic oscillator orbits underlies everything, especially the angular-momentum-carrying nature of circular orbits and Laguerre-Gaussian modes \cite{Allen:1992orbital}.
Our exposition will be pedagogical, with the aim of making the material accessible to new entrants to the field, as well as giving new insight to more seasoned researchers.

It is tempting, but potentially misleading, to think of optical angular momentum of structured Gaussian beams directly in terms of three-dimensional quantum spin.
In paraxial beams (as considered here), there is only one possible direction of spin or orbital angular momentum, namely the propagation direction, whereas more general quantum spins may be turning about any axis in three dimensions.
The analogy instead lies with the structure of different bases of representation.
For polarization, this is any pair of orthogonal elliptic polarization states (e.g.~linear horizontal and vertical, or right- and left-handed circular polarizations), and these basis states are parametrized by the Poincar{\'e} sphere \cite{bw:optics}.
This sphere is analogous to the Bloch sphere for quantum spin 1/2, as for light beams with a fixed direction of propagation, the electric field must be transverse, and hence a complex superposition of left and right circular polarizations, or equivalently, vibrations in the $x$ and $y$ directions.
Any pair of orthogonal polarizations (i.e.~complex 2-dimensional Jones vectors) are antipodal on the Poincar\'e sphere, as we will discuss in Section \ref{sec:poincare}.

For orbital angular momentum, this will be described using bases of Gaussian laser modes---especially the HG and LG basis sets---whose linear relationship is similar to quantum spin bases with different directions of rotation.
All of the discussion of Gaussian modes will be restricted to their amplitude distribution in the focal plane ($z=0$), so a fundamental Gaussian beam has amplitude $(2/\pi)^{1/2} w_0^{-1} \exp(-[x^2+y^2]/w_0^2)$, where $w_0$ represents the waist width of the beam \cite{siegman:lasers}, and is normalized (its square, integrated over the plane, gives unity).
This Gaussian has the same functional form as the ground state of a 2D quantum harmonic oscillator, which is justified physically \cite{siegman:lasers,Feldmann:1971modes,Nienhuis:1993paraxial,Habraken:2008orbital} in terms of the curved mirrors in the laser cavity having the effect on the paraxially-propagating wave, in the focal plane, of a harmonic potential.

\begin{figure}
\centering\includegraphics[width=\textwidth]{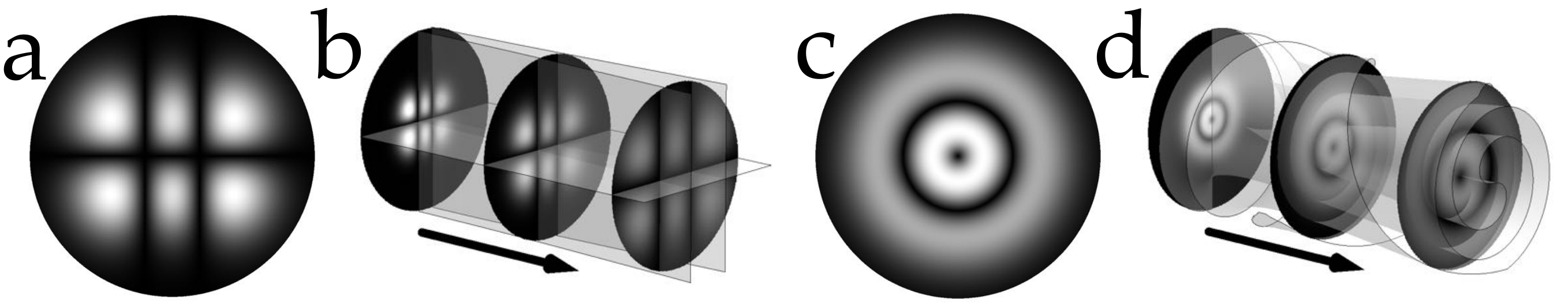}
    \caption{Illustrations of LG and HG beams in the focal plane, and upon propagation.
    (a) Intensity of $\HG_{21}$ in the focal plane. 
    (b) As the $\HG_{21}$ beam propagates, it spreads while maintaining the intensity pattern.  
    The orthogonal sheets represent the zeros of the HG beam, which are always at the positions of the zeros of the scaled Hermite polynomials.
    (c) Intensity of $\LG_{11}$ in the focal plane. 
    (d) As the $\LG_{11}$ beam propagates from its focal plane, it spreads maintaining the intensity pattern, but now each equiphase surface, such as the one represented by the grey surface, swirls around on propagation.
     }
    \label{fig:prop}
\end{figure}

Most laser cavities have residual astigmatism, breaking the cavity's pure axial symmetry; the HG modes (TEM modes) are higher-order modes of such cavities, given by
\begin{equation}
   \HG_{mn}(x,y) 
   = \frac{1}{w_0\sqrt{2^{m+n-1}\pi m!n!}} \exp\left(-\frac{x^2+y^2}{w_0^2}\right) H_m\left(\frac{\sqrt{2}x}{w_0}\right) H_n\left(\frac{\sqrt{2}y}{w_0}\right)
   \label{eq:hg}
\end{equation}
where $H_m, H_n$ denote Hermite polynomials \cite{lebedev:special}, and $m,n$ are nonnegative integers $0,1,2,\ldots$ (the fundamental Gaussian being the case $m = n = 0$).
HG modes are characterized by a nodal `grid' as seen in Figure \ref{fig:prop} (a), (b), and the function can be completely factorized into two functions, one depending on $x$ (indexed by $m$) and one on $y$ (indexed by $n$), and $N = m+n$ is the \emph{mode order}.
For mode order $N$, there are $N+1$ modes, for which $(m,n) = (N,0), (N-1,1), \ldots, (0,N)$.
The set of $\HG_{mn}$ modes is orthonormal (with respect to integration over the plane with uniform weight), and on propagation the modes maintain the same intensity pattern, even to the far field: the Fourier transform of an HG beam is functionally the same as (\ref{eq:hg}).
On propagation, the modes acquire an $N+1$-dependent Gouy phase factor; the intensity pattern of superpositions of modes with different $N$ changes on propagation, although it does not for superpositions with the same $N$.

The study of optical orbital angular momentum is mainly based around the LG modes \cite{Allen:1992orbital}, which are expressed in plane polar coordinates $R,\phi$ in the waist plane as
\begin{equation}
   \LG_{\ell p}(R,\phi) = \sqrt{\frac{2^{1+|\ell|} p!}{\pi (p+|\ell|!)}}\frac{1}{w_0^{1+|\ell|}}\exp\left(-\frac{R^2}{w_0^2}\right) R^{|\ell|} \exp(\rmi \ell \phi) L_p^{|\ell|}\left(\frac{2R^2}{w_0^2}\right),
   \label{eq:lg}
\end{equation}
whose radial dependence is determined by the associated Laguerre polynomial $L_p^{|\ell|}$ \cite{lebedev:special}, $p = 0,1,2,\ldots$ and $\ell$ is a positive or negative integer $0, \pm 1, \pm2,\ldots$.
LG modes factorize into an $R$-dependent function times $\exp(\rmi \ell \phi)$; this latter is an eigenfunction of the orbital angular momentum operator $-\rmi \partial_{\phi} = -\rmi(x \partial_y - y \partial_x)$ with eigenvalue $\ell$.
LG modes also occur as modes of laser cavities with mirrors with nonnegligible spherical aberration; due to residual astigmatism or localized cavity defects, there is a coupling of both signs of angular momentum, so the resulting LG cavity modes occur as the real and imaginary parts of (\ref{eq:lg}), and so do not carry a sense of right-handed or left-handed angular momentum.

LG modes also form a complete basis, have the same functional form as their Fourier transform, and maintain their intensity pattern on propagation, as shown for the example $\LG_{11}$ in Fig.~\ref{fig:prop}(c), (d). 
The phase swirls in such a way that the $z$-component of orbital angular momentum of the transverse beam is preserved, in addition to the Gouy phase.
An LG mode has mode order $N = |\ell|+2p$: for each $N$ there are again $N+1$ modes where $(\ell,p) = (-N,0), (-N+2,1), \ldots (N - 2,1), (+N,0)$.
Each LG mode with mode order $N$ can be expressed as a superposition of the $N+1$ HG modes of the same mode order, and vice versa \cite{Danakas:1992analogies}; a major aim of this paper is to explore this connection in detail.

Padgett and Courtial \cite{Padgett:1999poincare} observed that any superposition of Gaussian beams of mode order $N = 1$ can be represented on a sphere, analogous to the Poincar\'e sphere of polarization, as represented in Figure \ref{fig:poincare} (b): the circular modes $\LG_{\pm 1,0}$ occur at the poles, and linear modes $\HG_{10}$, $\HG_{01}$ with any orientation of Cartesian axes occur around the equator; intermediate `elliptic' states occur at other latitudes of the sphere.
It is natural to ask what happens to this sphere for different mode orders, and what is the connection with quantum spin and the Bloch sphere \cite{miles}.

In answering this question, we will make much of the 2-dimensional isotropic harmonic oscillator, using both classical Hamiltonian mechanics and its canonical quantization in quantum mechanics.
Since all classical and quantum properties of this system are well-known, we simply have to identify these with the known properties of elliptic polarization and Gaussian beams.
It is important to note that in this work, the classical picture is that of elliptic `rays' parametrized by the Poincar\'e sphere, and the wave mechanical one the Gaussian beams (as eigenfunctions of certain natural operators).
Although the language (and indeed the two-dimensional harmonic oscillator system) is suggestive of photons and quantum optics, all our classical, semiclassical and quantum discussion is between rays and waves.

In the next section, we will consider the Stokes parameters and Poincar\'e sphere, discussing historical approaches and the formulation in terms of the Hamiltonian mechanics of a 2D harmonic oscillator.
LG, HG and GG modes are considered in Section \ref{sec:quantum} as the eigenstates of operators naturally arising from the canonical quantization of the oscillator, and the connection is strengthened in Section \ref{sec:semic}, in which a semiclassical picture relating the `classical' polarization sphere and the `quantum' Gaussian mode sphere is described.
Throughout, the rectilinear `swinging' of linear polarization (and HG modes) contrasts with the `roundabout' motion of circular polarization (and LG modes).

\section{Elliptic polarization: Stokes, Poincar{\'e}, Gibbs and Hamilton}\label{sec:poincare} %% and Hamilton?

Poincar\'e himself gave a succinct description of the Poincar\'e sphere, in \cite{poincare:lumiere}, par 157 p284,
\begin{quotation}
\noindent The two poles ... correspond to the circular vibrations, and the various points of the first meridian [i.e.~the $s_1$-$s_3$ great circle] correspond to the ellipses whose axes are along the coordinate axes.
These ellipses are right[-handed] in the northern hemisphere, and left[-handed] in the southern hemisphere.
The orientation of the axes depends only on the longitude [on the sphere]; the loci of points [of the same axis alignment] ... are the [great] circles passing through the two poles, that is to say, the meridians.
The shape of the ellipse depends only on the latitude; the locus of the points corresponding to a given shape is a line of latitude.
\end{quotation} 
The sphere is a representation of the complex 2D Jones vector $\bs{E} = (E_x,E_y)$; in the circular basis, $\left( \begin{smallmatrix} E_{\mr{RH}} \\ E_{\mr{LH}}\end{smallmatrix}\right) = 2^{-1/2}\left(\begin{smallmatrix} 1_{\phantom{.}} & -\rmi \\ 1_{\phantom{.}} & \rmi \end{smallmatrix}\right)\left(\begin{smallmatrix} E_x \\ E_y \end{smallmatrix}\right)$.
Poincar\'e effectively constructed the sphere by stereographically projecting $E_{\mr{LH}}/E_{\mr{RH}} =\tan(\theta/2)\exp(\rmi \phi),$ where $\theta,\phi$ are spherical angles $0 \le \theta \le \pi,$ and $0\le 0 \le \phi < 2\pi$.

It is usually more convenient to define the Poincar\'e sphere using the \emph{Stokes parameters}
\begin{equation}
   \left.
   \begin{array}{cclcccl}
   S_0 & = & I & = & |E_x|^2 + |E_y|^2 & = & \bs{E}^*\cdot\bs{\sigma}_0\cdot\bs{E}, \\
   S_1 & = & I_{0^{\circ}} - I_{90^{\circ}} & = & |E_x|^2 - |E_y|^2 & = & \bs{E}^*\cdot\bs{\sigma}_3\cdot\bs{E}, \\
   S_2 & = & I_{45^{\circ}} - I_{135^{\circ}} & = & E_x^* E_y + E_x E_y^* & = & \bs{E}^*\cdot\bs{\sigma}_1\cdot\bs{E}, \\
   S_3 & = & I_{\mr{RH}} - I_{\mr{LH}} & = & -\rmi (E_x^* E_y - E_x E_y^*) & = & \bs{E}^*\cdot\bs{\sigma}_2\cdot\bs{E}.
   \end{array}\right\}
   \label{eq:stokes}
\end{equation}
The first column of equalities indicates how the Stokes parameters can be found by intensity measurements, with polarizer at angle given in the subscript ($0^{\circ}, 90^{\circ}, 45^{\circ}, 135^{\circ}$, and $\mr{RH},\mr{LH}$ circular polarizations).
The second column gives the Stokes parameters in terms of components of $\bs{E}$, which in fact are complex inner products of $\bs{E}$ with the Pauli spin matrices $\bs{\sigma}_0 = \left( \begin{smallmatrix} 1 & 0 \\ 0 & 1 \end{smallmatrix}\right), \bs{\sigma}_1 = \left( \begin{smallmatrix} 0 & 1 \\ 1 & 0 \end{smallmatrix}\right), \bs{\sigma}_2 = \left( \begin{smallmatrix} 0 & -\rmi \\ \rmi & 0 \end{smallmatrix}\right), \bs{\sigma}_3 = \left( \begin{smallmatrix} 1 & 0 \\ 0 & -1 \end{smallmatrix}\right)$; the subscripts on the $S_i$ agree with the subscripts on $\bs{\sigma}_i$ in the circular basis.
They satisfy $S_1^2 + S_2^2 +S_3^2 = S_0^2$, so we can define the normalized \emph{Stokes vector} $\bs{s} = (s_1,s_2,s_3)$, where $s_i = S_i/S_0$, $i = 1,2,3$, and all possible choices of $\bs{s}$ define a unit sphere.
Multiplying $\bs{E}$ by a complex phase factor (representing, for instance, the evolution of time) does not change $\bs{s}$.

The fact that the Pauli matrices appear in this classical setting suggests that the physics of polarization may resemble that of quantum spin; if $\bs{E}$ were a quantum state of spin 1/2---rather than a classical state of light---the Stokes vector $\bs{s}$ would represent the axis of rotation of the quantum spin state on the Bloch sphere.
However, 3D Stokes space does not directly correspond to real space (e.g.~Stokes vectors $(1,0,0)$ and $(-1,0,0)$ are $180^{\circ}$ apart in Stokes space, but represent orthogonal polarizations $(1,0), (0,1)$).

Although R C Jones is credited with developing the Jones calculus of 2D vectors and matrices in 1943 \cite{Jones:1943calculus}, it was apparently J W Gibbs, in his original work on vectors privately published in 1884, who first identified $\bs{E} = (E_x,E_y)$---which Gibbs called a `bivector'--- with a `directional ellipse' \cite{gibbs:vector}. 
(Most of the content of Gibbs' notes also appears in the final chapter of the book published later by Wilson \cite{wilson:gibbs}, with explicit application to elliptically polarized light.)
%%% (see also \cite{??gibbsbook}).) -- what was this?

\begin{figure}
\centering\includegraphics[width=4.5cm]{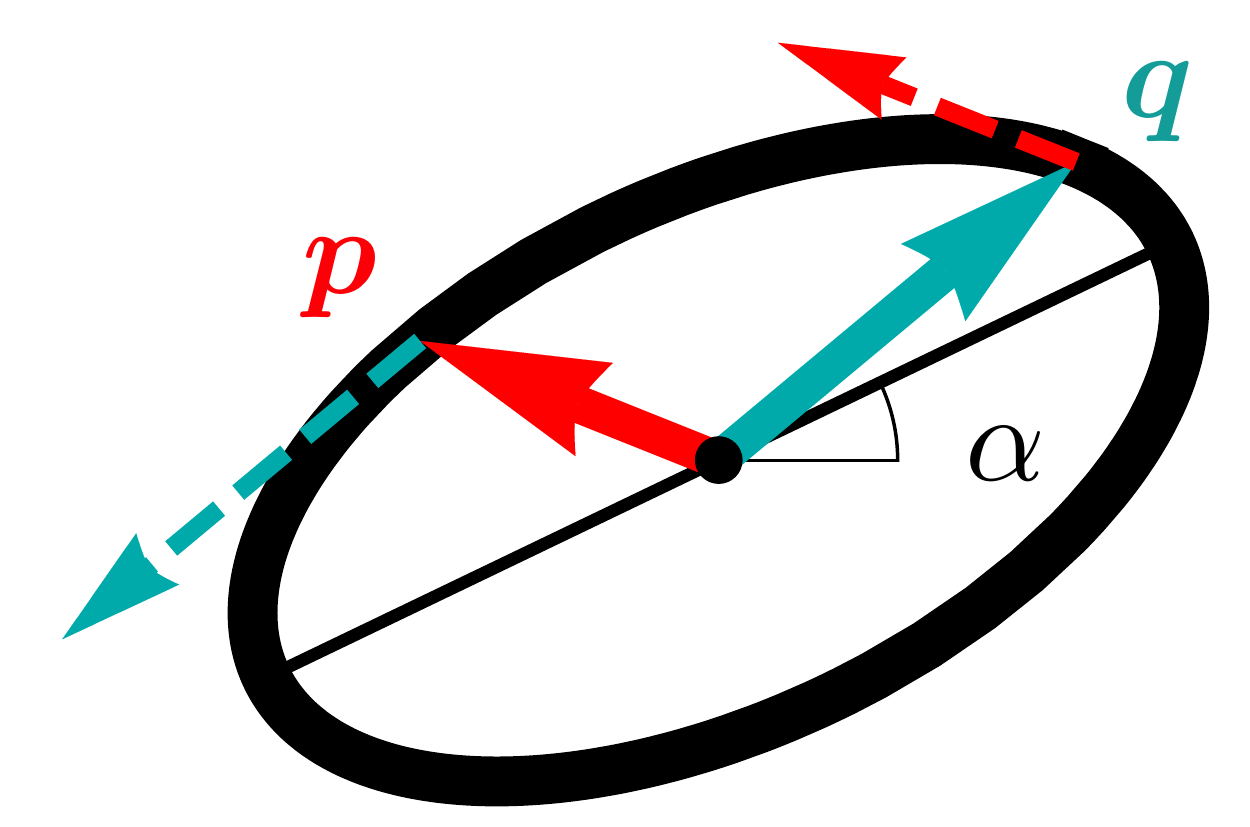}
    \caption{Gibbs' ellipse `bivector' construction for $\bs{E} = \bs{q}+\rmi\bs{p}$.
    Both $\bs{q}$ and $\bs{p}$ lie on the ellipse, such that the velocity at $\bs{q}$ is parallel to $\bs{p}$, and the velocity at $\bs{p}$ is parallel to $-\bs{q}$.
    The angle of the major axis with the $x$-axis is $\alpha$.
     }
    \label{fig:gibbs}
\end{figure}

Gibbs wrote $\bs{E} \equiv  \bs{q} + \rmi  \bs{p},$ in terms of its real and imaginary parts, noting that multiplication by a phase factor should not alter the geometry of the complex vector.
For our monochromatic polarization state (for simplicity with $\omega = 1$), the physical, time-dependent electric vector is 
\begin{equation}
   \bs{Q}(t) = \re[\bs{E}\exp(-\rmi t)] = \bs{q} \cos t + \bs{p} \sin t.
   \label{eq:Qdef}
\end{equation}
At $t = 0$, $\bs{Q}(0) = \bs{q}$, and its velocity (i.e.~instantaneous linear momentum) is $\dot{\bs{Q}}(0) = \bs{p}$ (hence our choice of $\bs{q} + \rmi \bs{p}$, rather than the often-seen $\bs{p}+\rmi \bs{q}$ \cite{bw:optics}).
By elementary geometry, $\bs{Q}(t)$ traces out an ellipse as $0 \le t \le 2\pi$.
This construction is shown in Fig.~\ref{fig:gibbs}: the ellipse has a handedness (orientation) defined by the right-handed or left-handed rotation of the vector moving on the shorter arc from $\bs{q}$ to $\bs{p}.$
For any choice of $t$, the counterpart $\bs{P}(t)$ of $\bs{Q}(t)$ analogous to $\bs{p}$ to $\bs{q}$ is
\begin{equation}
   \bs{P}(t) = \im[\bs{E}\exp(-\rmi t)] = \bs{p} \cos t - \bs{q} \sin t.
   \label{eq:Pdef}
\end{equation}

The Stokes parameters can be rewritten in terms of $\bs{q}$ and $\bs{p}$, or equivalently $\bs{Q}$ and $\bs{P}$,
\begin{equation}
   \left.
   \begin{array}{cclcl}
   S_0 & = & p_x^2 + q_x^2 + p_y^2 + q_y^2 & = & P_x^2 + Q_x^2 + P_y^2 + Q_y^2,  \\
   S_1 & = & p_x^2 + q_x^2 - p_y^2 - q_y^2 & = & P_x^2 + Q_x^2 - P_y^2 - Q_y^2, \\
   S_2 & = & 2(p_x p_y + q_x q_y) & = & 2(P_x P_y + Q_x Q_y), \\
   S_3 & = & 2(q_x p_y - q_y p_x) & = & 2(Q_x P_y - Q_y P_x).
   \end{array}\right\}
   \label{eq:stokespq}
\end{equation}
In this form, the Stokes parameters are quantities in classical \emph{Hamiltonian mechanics} with position $\bs{Q}$ and momentum $\bs{P}$: the electric vector's motion is that of a 2D, isotropic harmonic oscillator with Hamiltonian $H = \tfrac{1}{2}S_0$, which is conserved in time -- this agrees with a mechanical harmonic oscillator with mass and spring constant (and angular frequency) set to unity. %%action angle variables
The second column of equalities in (\ref{eq:stokespq}) mean that the Stokes parameters are all formally \emph{constants of the motion}; the angular momentum $L = \tfrac{1}{2}S_3$, and we define $M$ as the difference of the Hamiltonians in $x$ and $y$, i.e.~$M \equiv H_x - H_y = \tfrac{1}{2} S_1$; the difference of the Hamiltonians at $45^{\circ}$ and $135^{\circ}$ is $\overline{M} \equiv H_{45^{\circ}} - H_{135^{\circ}} = \tfrac{1}{2} S_2$.

The Hamiltonian dynamics of the 2D harmonic oscillator is extremely well understood \cite{goldstein:mechanics,Fradkin:1967existence,holm:geometricI}, and much follows from consideration of the Poisson bracket
\begin{equation}
   \{ A, B\} = \frac{\partial A}{\partial \bs{q}}\cdot \frac{\partial B}{\partial \bs{p}} - \frac{\partial B}{\partial \bs{q}}\cdot \frac{\partial A}{\partial \bs{p}},
   \label{eq:poisson}
\end{equation}
where $\partial/\partial\bs{q} = (\partial_{q_x},\partial_{q_y})$, etc.
The Poisson brackets of $H,L,M,\overline{M}$ satisfy 
\begin{equation}
   \{ L,H\} = \{ M,H\} = \{ \overline{M},H\} = 0, \qquad \{ L, M\} = 2\overline{M}, \qquad \{ M, \overline{M}\} = 2L, \qquad \{ \overline{M}, L\} = 2M.
   \label{eq:poissonalg}
\end{equation}
These Poisson commutation relations between $L,M,\overline{M}$ therefore form an $\rm{su}(2)$ algebra \cite{goldstein:mechanics,holm:geometricI}, exactly the same algebraic structure between Cartesian components of 3D angular momentum, $L_x, L_y, L_z$.
It reveals a deep connection between the 2D harmonic oscillator and 3D rotation, in particular accounting for the occurrence of the Pauli spin matrices in (\ref{eq:stokes}), without directly invoking quantum mechanics, and anticipates many properties of the 2D quantum harmonic oscillator, such as the Schwinger oscillator representation (as discussed below).

A given state of elliptic polarization has specific values for each of $L, M, \overline{M}$ as well as $H$ -- being $\tfrac{1}{2}$ times the four Stokes parameters, they satisfy $L^2 + M^2 + \overline{M}^2 = H^2$.
In fact, any unit vector $\bs{u} = (X,Y,Z)$, corresponds to a constant of the motion by the inner product $C = \bs{u}\cdot(M,\overline{M},L)$.
For the polarization state whose Stokes vector $\bs{s} = \bs{u}$, $C = H$, and antipodally, $\bs{s} = -\bs{u}$, $C = -H$; the basis of $L,M,\overline{M}$ of the $\rm{su}(2)$ Poisson algebra is convenient, but no more unique than the choice of linear and circular polarization states in defining the Stokes parameters.

The Hamiltonian formulation gives an insight into polarization geometry in terms of abstract phase space, which is 4-dimensional with coordinates $(Q_x,Q_y,P_x,P_y)$, and initial coordinates $(q_x,q_y,p_x,p_y)$.
All polarization vectors with the same intensity $S_0 = 2H$, that is the same value of the Hamiltonian, lie on a 3-sphere in 4-dimensional phase space, $Q_x^2 + Q_y^2+P_x^2 +P_y^2 = H^2$ constant.
The elliptic polarizations are all periodic orbits which are topologically circles: the 3-sphere is cut up (technically \emph{fibred}) by these loops.
The space which parametrizes these loops is of course the Poincar\'e sphere, where each point on the sphere $\bs{s} = (s_1,s_2,s_3)$ corresponds to a polarization state, that is, a loop in the 3-sphere.
In fact, this is the celebrated \emph{Hopf fibration} \cite{urbantke:2003hopf}, by which the 3-sphere is fibred by 1-spheres (circles) with base space the 2-sphere; the Stokes parameters (\ref{eq:stokespq}) define a standard form for the Hopf map \cite{Dennis:2002polarization}.
In this 3-sphere, a contour of constant $S_1$ (or $S_2$ or $S_3$) defines a 2-torus: fixing the value of all three (respecting $S_0^2 = S_1^2 + S_2^2 + S_3^2$) defines a single orbit along which three tori intersect.
The rich structure of this Hamiltonian system comes from the fact that it is superintegrable \cite{Kalnins:1996superintegrability}, with more independent constants of the motion (three) than there are classical degrees of freedom (two, the dimension of configuration space); there are even more symmetries in the 3D harmonic oscillator \cite{Fradkin:1965three,Fradkin:1967existence}.

A feature of integrable systems is that they admit a reparametrization, by a suitable canonical transformation, to action-angle variable (whose generalized position coordinate is an angle $\chi$, and generalized momentum $I$ a constant of the motion).
In the 2D isotropic harmonic oscillator, one set of action-angle coordinates is $\chi = t$ ($0 \le t \le 2\pi$), and $I = H = \tfrac{1}{2} S_0$.
Another pair of action-angle variables defines the particular elliptical orbit: the conjugate angular variable to angular momentum $L$ is $\alpha$, the angle between the $x$-axis and the major axis of the ellipse (Figure \ref{fig:gibbs}).
$\alpha$ is half the azimuthal angle about the vertical axis of the Poincar\'e sphere,
\begin{equation}
   \alpha = \tfrac{1}{2}\arg(M + \rmi \overline{M}) = \tfrac{1}{2}\arg(S_1 + \rmi S_2).
   \label{eq:alpha}
\end{equation}
It is straightforward but tedious to show that the Poisson commutator satisfies $\{ \alpha,L\} = 1$, appropriate for conjugate variables.
Rather than defining the polarization state by $\bs{q}$ and $\bs{p}$, we can use the action-angle pairs $H,t$ and $L, \alpha$.
The ellipse orientation $\alpha$ is the harmonic oscillator analogue to the Runge-Lenz vector for Keplerian orbits \cite{goldstein:mechanics,Fradkin:1967existence}.
The angles around the other axes of the sphere are conjugates to the other constants of the motion $M$, $\overline{M}$, and any choice of pair  parametrizes the state of polarization.

Despite the apparent equal footing of $L, M, \overline{M}$ as dynamical quantities, $L$ is fundamentally different from $M$ and $\overline{M}$.
The momentum terms in $M$ only occur quadratically, such as $p_x^2$, and therefore $M$ takes the same value regardless of the sign of transverse momentum.
This is of course necessary for a constant of swinging motion of the linear orbits of a 2D harmonic oscillator, represented by a point on the equator of the Poincar\'e sphere (and in which opposite directions in real space are identified).
This contrasts with angular momentum $L$, which has a definite sign of rotational motion---a roundabout---that depends on the relative phase between swings in $x$ and $y$.
From the Hamiltonian viewpoint, this distinction comes from the fact that although $M, \overline{M}$ and $L$ are all quadratic in the canonical variables $Q_x,Q_y,P_x,P_y$ (and hence on an equal footing in the Poisson algebra), $L$ is linear in the momentum variables alone (also the position variables alone), whereas $M$, $\overline{M}$ are quadratic in the momenta, and also the positions; the constants of the motion can detect the sign of angular momentum, but not linear momentum.

Of course, this set up can be quantized directly to study the quantum polarization properties of photons.
However, we instead are considering a different physical situation, that of the scalar behaviour of Gaussian laser modes in the focal plane, which nevertheless has strong mathematical analogies with this view of the Poincar\'e sphere, even if the physical relationship is not (yet) clear.

\section{HG and LG modes as eigenfunctions of the 2D quantum harmonic oscillator}\label{sec:quantum}%% cite: Steuernagel somehwere?

We have already stated that laser cavity modes in the focal plane are analogous to 2D quantum harmonic oscillator eigenfunctions.
In fact, the HG modes (\ref{eq:hg}) are exactly 2D quantum harmonic oscillator eigenstates (albeit with the convention that the usual width of a quantum oscillator $w = w_0/\sqrt{2}$ for $w_0$ the optical focal beam waist); as in the previous section, we set our constants to be dimensionless, and the appropriate choice is $w = 1, w_0 = 1/\sqrt{2}$.
Having introduced the classical Hamiltonian formalism in the previous section, we can observe that the wave mechanics picture (i.e.~``quantum mechanics'') comes purely from canonical quantization, where observables $H, L, M$ and $\overline{M}$ become operators on wavefunctions, and in configuration space, $\boldsymbol{p} \to \widehat{\boldsymbol{p}} = -\rmi (\partial_x, \partial_y)$ (in our optical wave mechanics, we have no $\hbar$). 
We defer the \emph{meaning} of the quantization procedure (and, indeed, the significance to Gaussian beams of the classical harmonic oscillator) until later  -- the fact that HG and LG modes \emph{are} eigenstates of the 2D quantum isotropic harmonic oscillator should be sufficient motivation to continue for now \cite{rays}.

The 2D harmonic oscillator is separable in Cartesian coordinates, since the quantum Hamiltonian $\widehat{H} = \tfrac{1}{2}(\widehat{\bs{p}}^2 + \widehat{\bs{q}}^2)$ commutes with the Hamiltonian in $x$, $\widehat{H}_x = \tfrac{1}{2}(\widehat{p}_x^2 + \widehat{q}_x^2)$. 
Therefore there are eigenfunctions of $\widehat{H}$ which factorise into functions of $x$ and $y$ (each 1D harmonic oscillator eigenfunctions), and of course these are the HG modes, for which (with our choice of dimensionless constants) $\widehat{H}_x \HG_{mn} = (m + \tfrac{1}{2}) \HG_{mn}$, $\widehat{H}_y \HG_{mn} = (n + \tfrac{1}{2}) \HG_{mn}$. % setting constants to unity
Thus 
$$\widehat{H}\HG_{mn} = (\widehat{H}_x + \widehat{H}_y)\HG_{mn} = (m + n + 1)\HG_{mn} = (N+1) \HG_{mn},$$
so the mode order $N$ corresponds to the quantum harmonic oscillator's principal quantum number, and the quantum energy eigenvalue $N+1$ is related to the total Gouy phase on propagation \cite{Nienhuis:1993paraxial}.

In fact, we can consider the operator analogue of the classical quantity $M$,
\begin{equation}
   \widehat{M} = \widehat{H}_x - \widehat{H}_y = \tfrac{1}{2}(\widehat{p}_x^2 + \widehat{q}_x^2 - \widehat{p}_y^2 - \widehat{q}_y^2).
   \label{eq:Mhat}
\end{equation}
The commutator $[\widehat{M},\widehat{H}] = 0$ (as we could have anticipated from the Poisson brackets of the classical observables), and so there is a mode set of simultaneous eigenfunctions of $\widehat{M}$ and $\widehat{H}$ -- this is the HG modes.
In fact, $\widehat{M}\HG_{mn} = (m - n)\HG_{mn}$: the spectrum of $\widehat{M}$ consists of integers $-N, -N+2, \ldots, N$; this ladder spectrum resembles that of an angular momentum operator (with an extra factor of 2), originating from isomorphism of the underlying $\rm{su}(2)$ Poisson algebras.
$\widehat{M}$ is the difference of two Hamiltonians, distinguishing a set of Cartesian axes (up to a rotation by $180^{\circ}$), but no sense of linear or angular momentum associated with it.

The $\LG$ modes are of course eigenfunctions of angular momentum, with $\widehat{L} = \widehat{q}_x\widehat{p}_y-\widehat{q}_y\widehat{p}_x = \rmi \partial_{\phi}$, and despite having different eigenfunctions, the spectrum of $\widehat{L}$, being the allowed values $\ell$, i.e.~$-N, -N+2, \ldots, N-2, N$, is identical to that of $\widehat{M}$.
The steps of 2 are required with each unit increment in $p$, since $N = |\ell| + 2p$. 
Therefore, with our definition, even mode orders only consist of modes with even angular momentum, and likewise for odd.
Mathematically, this can be understood from the analytic form of the LG modes: (\ref{eq:lg}) must be an analytic function of its arguments $R$ and $\phi$ (equivalently, $x$ and $y$) everywhere in the transverse plane, including at the origin, and the mode order $N$ is the maximum power of $R$ multiplying the Gaussian (equal to the sum of the degrees $m +n$ of the Hermite polynomials).
The argument of the Laguerre polynomial must be proportional to $R^2$, since a polynomial with odd powers in $R =\sqrt{x^2+y^2}$ is not analytic at the origin, and increasing the polynomial degree by one ($p \to p+1$) changes the overall degree of the polynomial factor by 2, which must be compensated by an extra factor $R^2 \exp(\pm 2\rmi \phi)$.
This argument justifying the separation between modes of even and odd azimuthal order holds for other modes expressed naturally in polar coordinates, such as Zernike functions used as a basis in a circular pupil \cite{bw:optics}.

The canonical quantization procedure defines three operators $\widehat{L}, \widehat{M}, \widehat{\overline{M}}$ as counterparts to the classical observables $L, M, \overline{M}$; although we have not discussed the third of these, clearly its eigenfunctions are Hermite-Gaussian modes with Cartesian axes oriented at $45^{\circ}$ and $135^{\circ}$ to the $x$-axis.
As canonical quantization is based on the canonical commutation relations $[\widehat{q}_x,\widehat{p}_x] = [\widehat{q}_y,\widehat{p}_y] = \rmi$ (other commutators zero), the commutation relations between $\widehat{L}, \widehat{M}, \widehat{\overline{M}}$ is $[\widehat{L},\widehat{M}] = 2\rmi \widehat{\overline{M}},$ etc., equivalent to the Poisson bracket relations between their classical counterparts.
This $\rm{su}(2)$ operator algebra is mathematically analogous to the 3D quantum angular momentum relations between $\widehat{L}_x, \widehat{L}_y, \widehat{L}_z$.
The connection has been very important historically, and was used by Schwinger to derive properties of quantum angular momentum (such as Clebsch-Gordan coefficients) from known behaviour of quantum harmonic oscillators \cite{schwinger:AM,sakurai:qm}.
This connection was also discussed explicitly by Danakas and Aravind in \cite{Danakas:1992analogies}, who described in detail the connection between HG and LG laser modes and the 2D quantum oscillator, although without the connections with the classical analogue.

\begin{figure}
\centering\includegraphics[width=10cm]{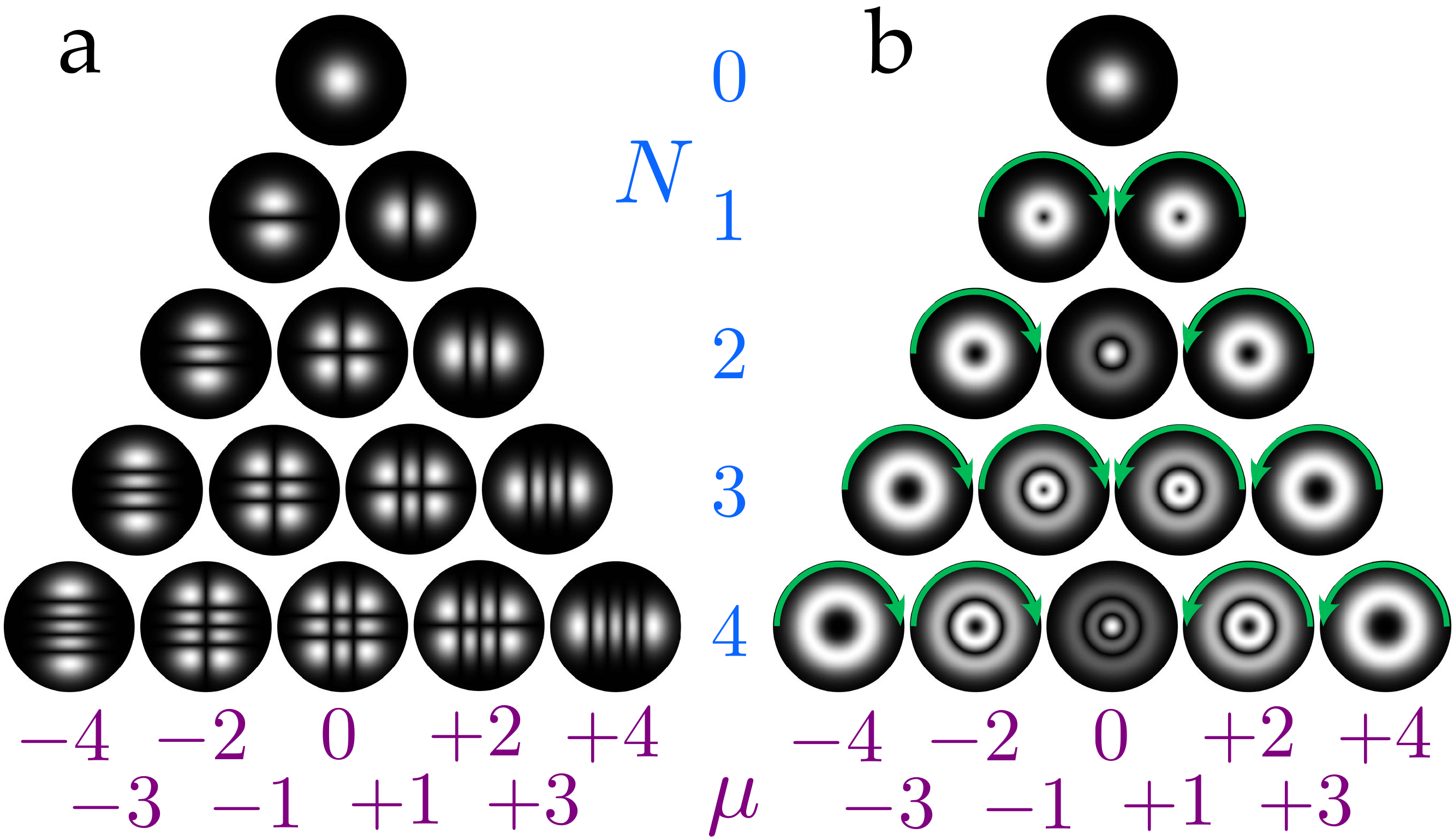}
    \caption{Gaussian mode sets as eigenfunctions with eigenvalues $\mu$ and $N$, for $N = 0,1,2,3,4$.
    (a) HG modes; (b) LG modes.
     }
    \label{fig:modes}
\end{figure}

LG and HG modes are therefore eigenfunctions of the operators $\widehat{L}, \widehat{M}$, and HG modes rotated by $45^{\circ}$ are eigenfunctions of $\widehat{\overline{M}}$.
These operators all have the same spectrum, labelled by $\mu = -N, -N+2, \ldots,N-2, N$ for mode order $N$.
Each is a basis for Gaussian modes of mode order $N$, inequivalent as the commutator of the corresponding operators does not vanish.
The lowest order members of the HG and LG sets, labelled by $N$ and $\mu$, are shown in Figure \ref{fig:modes}.
The `swings and roundabouts' distinction between linear back-and-forth motion associated with $M$ and angular motion with a definite sign associated with $L$ gives a physical interpretation to the fact that HG modes are real valued, resembling a standing wave pattern, whereas LG modes are complex with a definite sense of angular momentum direction in the azimuthal phase factor $\exp(\rmi \ell \phi)$, although both HG and LG modes have the same eigenvalues.
Real LG modes such as those discussed by \cite{siegman:lasers}, with angular dependence $\cos \ell\phi$ or $\sin\ell \phi$, are not eigenfunctions of $\widehat{L}$, but are of $\widehat{L}^2$: they may occur in laser cavities whose mirrors have some spherical aberration \cite{siegman:lasers}, or if a defect in the cavity generates a nodal line.
Since we are considering stationary modes, it is natural to expect real LG modes occurring as modes of cavities, like HG modes; devices such as cylindrical lenses (or indeed, spatial light modulators) are required to synthesize LG beams carrying orbital angular momentum of definite sign, as appreciated by Allen \emph{et al.}~\cite{Allen:1992orbital}.

Just as alternative classical constants of the motion can be made by linear combinations, so can new operators such as $\widehat{C} = \bs{u}\cdot(\widehat{M},\widehat{\overline{M}},\widehat{L})$ involving the scalar product of the unit vector $\bs{u} = (X,Y,Z)$ with the vector of operators $(\widehat{M},\widehat{\overline{M}},\widehat{L})$.
$\widehat{C}$ is the canonical quantization of the constant of the motion $C = X M + Y \overline{M} + Z L$ corresponding to the elliptic polarization state at $\bs{s} = \bs{u}$ on the Poincar\'e sphere: for every choice of $\bs{u}$, $\widehat{C}$ has the same spectrum, $\mu = -N, \ldots N$.
If $\bs{u} = (\cos(2\alpha)\sin\beta,\sin(2\alpha)\sin\beta,\cos\beta)$, $\widehat{C}$ can be found from $\widehat{L}$ by appropriate rotations through angles $\beta = \theta$ and $2\alpha = \phi$.
Thus the eigenfunctions of $\widehat{C}$, the \emph{Generalized Hermite-Laguerre Gaussian beams} \cite{Abramochkin:2004generalized,Visser:2004oam,Dennis:2006rows} (GG beams) are superpositions of LG beams, with coefficients the appropriate matrix elements of the quantum mechanical rotation operator of the equivalent spin $N/2$ (with $\mu/2$ playing the role of $m$ in quantum angular momentum),
\begin{equation}
   \GG_{N,\mu,\alpha,\beta} = \rmi^{(\mu - N)/2} \sum_{\mu'/2 = -N/2}^{N/2} (-1)^{(\mu'-|\mu'|)/2} \exp(-\rmi \mu' \alpha) d_{\mu/2,\mu'/2}^{N/2}(\beta) \LG_{\mu',(N-|\mu'|)/2},
   \label{eq:GG}
\end{equation}
where $d^j_{m',m}(\beta)$ denotes the usual Wigner-$d$ function \cite{sakurai:qm}, and the extra factors of $-1$ and $\rmi$ are required to make the LG modes have the correct phases as the quantum spin states (i.e.~imposing the Condon-Shortley convention).
When $\alpha = 0, \beta = \pi/2$, $\GG_{N,\mu,0,\pi/2} = \HG_{(N+\mu)/2,(N-\mu)/2}$, as described by \cite{Danakas:1992analogies}.
Just as the HG modes can be thought of as the wave analogue of back-and-forth harmonic oscillator orbits, and LG modes the quantizations of circular orbits, these are the wave states corresponding to general elliptic polarization.
However, there are no natural coordinates to express a GG beam, for general $\beta$ -- they can only be expressed as sums like (\ref{eq:GG}) (or analogous sums of HG modes), with amplitudes weighted by quantum rotation matrix elements.
The GG beams can be created from LG or HG beams via rotated cylindrical lenses (the rotation angle giving $\beta$) in a similar way to birefringent wave plates providing rotations on the Poincar\'e sphere \cite{ONeil:2000mode} (effectively increasing the relative oscillation rate in one transverse direction with respect to the other), and sequences of modes forming closed loops on the sphere give rise to geometric phases, just like the Pancharatnam phase for polarization \cite{Galvez:2003geometric,Calvo:2005wigner}.

\begin{figure}
\centering\includegraphics[width=\textwidth]{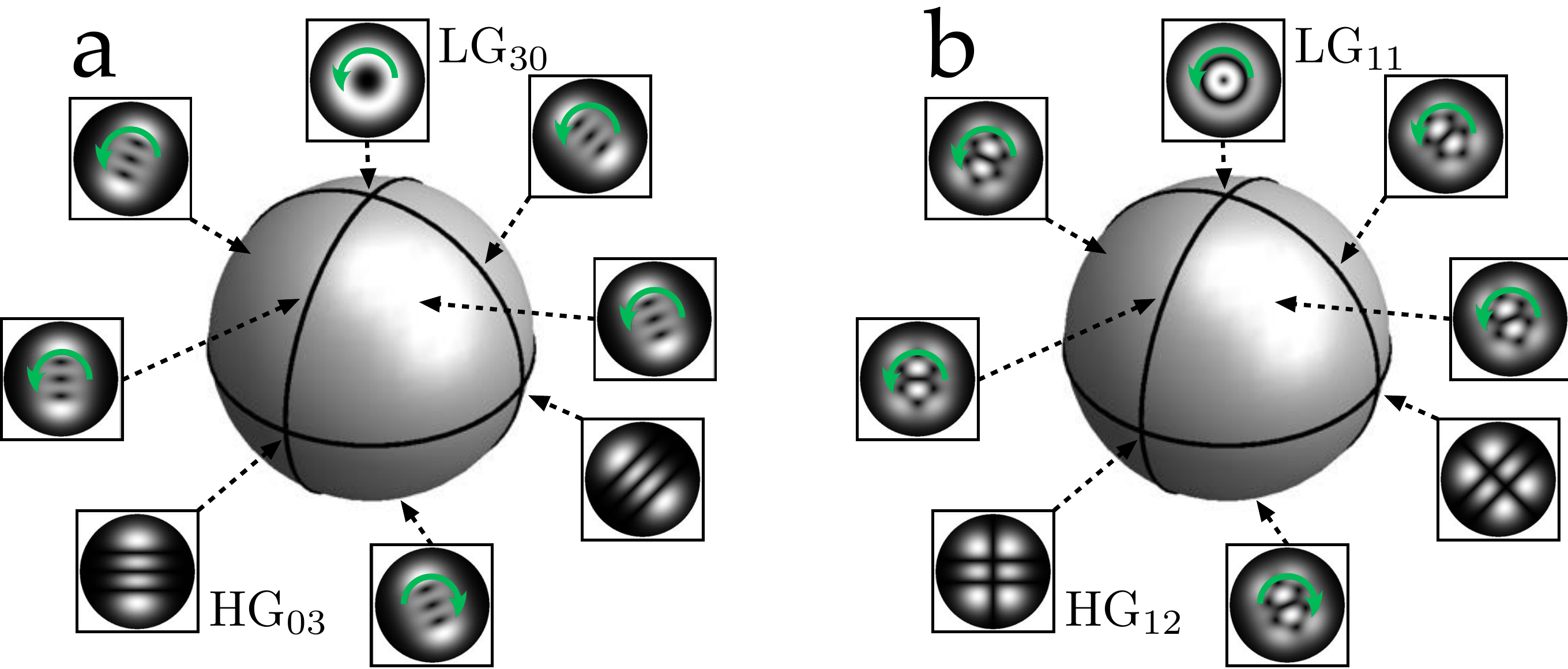}
    \caption{Gaussian mode spheres for $N = 3$. (a) $\mu = +3$; (b) $\mu = +1$.
     }
    \label{fig:poincare3}
\end{figure}

We can now understand the Poincar\'e sphere for Gaussian modes: we really have a family of operators parametrized by the points on a unit sphere, $X \widehat{M} + Y \widehat{\overline{M}} + Z \widehat{L}$, corresponding to the classical constants of the motion which are linear combinations of $L, M, \overline{M}$.
Each of these operators commutes with $\widehat{H}$, and has the same eigenvalues $\mu = -N, -N+2, \ldots, N$ regardless of the direction of $\bs{u}$.
For each choice of $N$ and $\mu$, there is therefore a family of eigenfunctions parametrized by points on the sphere, each of which is a Gaussian wave field corresponding to the classical oscillator state at the same point on the Poincar\'e sphere: LG modes at the poles corresponding to circular motion, (possibly rotated) HG modes around the equator corresponding to swinging motion, and other GG modes elsewhere, corresponding to elliptical oscillation.
There is an antipodal symmetry between the $\mu$-sphere at $(X,Y,Z)$ and the $-\mu$-sphere at $(-X,-Y,-Z)$, inherited from the orthogonality of antipodal polarizations of the Poincar\'e sphere. 
The original sphere of Padgett and Courtial (Figure \ref{fig:poincare} b) were the states with eigenvalues $N = 1, \mu = 1$, representing all the types of Gaussian modes (LG, HG and GG), which are all the Gaussian modes of mode order one.
For higher values of $N$ there are different spheres, one distinct for each value of $|\mu|$ \cite{habraken:thesis}, with examples for $N = 3$ shown in Figure \ref{fig:poincare3}.
This parametrization of different eigenfunctions with the same eigenvalue by points on the sphere is analogous to quantum spin states, oriented in different directions but with the same quantum numbers $j, m$, which are eigenfunctions of the component of the spin operator in different directions; in both cases, the general state is represented as a superposition of a different basis set with coefficients the Wigner $d$-functions (rotation matrix elements).
This Hermite-Laguerre sphere can also be constructed in terms of the raising and lowering operators of the harmonic oscillator \cite{Visser:2004oam}, which takes advantage of the $su(2)$ structure of the system, in the spirit of the Schwinger oscillator.

Following the formalism of canonical quantization of the classical 2D harmonic oscillator as discussed in the previous section, we have accounted for the Poincar\'e sphere-like structure of HG, LG and GG modes, explaining the connection in terms of operators and their eigenvalues and eigenfunctions, but without detailed physical justification of what the classical oscillator has to do with laser modes.
Before we do this, we explore the connection between the classical and quantum states further, by considering a hybrid, semiclassical picture which brings out further aspects of the connection.

\section{Semiclassical picture of the Poincar\'e sphere for Gaussian beams}\label{sec:semic}

Having now considered the Hamiltonian dynamics of the 2D isotropic oscillator underlying the Poincar\'e sphere, and its canonical quantization to give operators whose eigenfunctions are HG, LG and GG beams, we ask whether there is a geometric way of visualizing this connection, or is the quantum notion simply abstract?

Semiclassical methods, connecting the classical and quantum approaches, are exact for harmonic oscillators, for instance Bohr-Sommerfeld quantization according to the integral of a pair of action-angle variables $(\chi,I)$,
\begin{equation}
   \int_0^{2\pi} I \rmd\chi = 2\pi j,
   \label{eq:bohrsommerfeld}
\end{equation}
where $j$ is an integer (the right hand side should be multiplied by $\hbar$, but this is set to be unity here).
With $I = H$ the Hamiltonian and $\chi = t$, $j$ becomes the Hamiltonian eigenenergy $N+1$.
Semiclassical integrals of this form for other action-angle variables can be interpreted geometrically, where quantum states semiclassically are associated with paths on the classical Poincar\'e sphere of harmonic oscillator orbits by (\ref{eq:bohrsommerfeld}).

\begin{figure}
\centering\includegraphics[width=\textwidth]{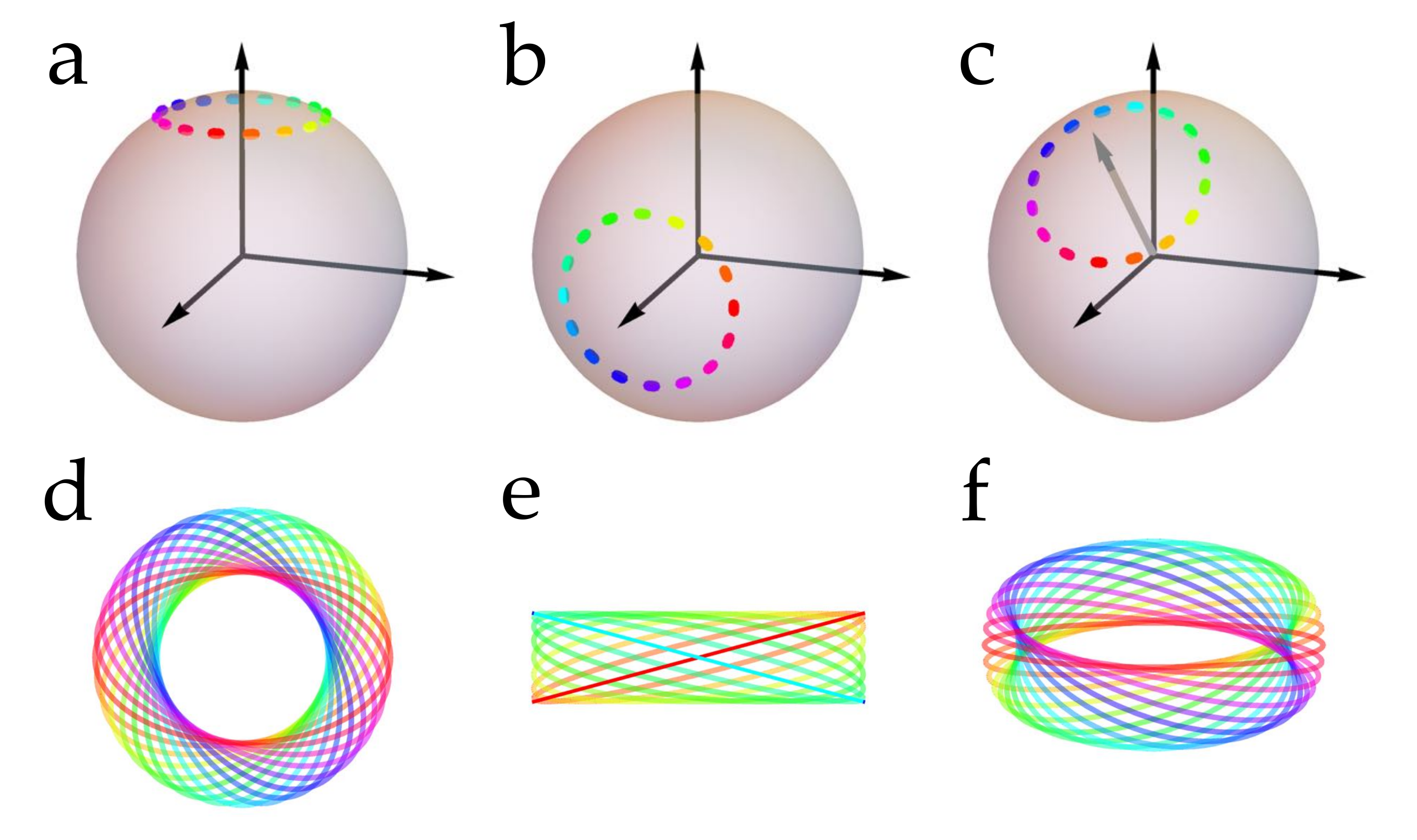}
    \caption{Families of ellipses with well-defined classical constants of motion, corresponding to wave/quantized states.
    (a)-(c) curves on the Poincar\'e sphere with fixed classical constant of motion, tracing out a circle with constant: (a) $Z$; (b) $X$; (c) $X+Z$.
    These correspond to families of ellipses which underlie the LG, HG and GG beams respectively: (a) family with circular caustics; (b) family with rectangular caustics; (c) family with more general caustics.  
     }
    \label{fig:families}
\end{figure}

An energy eigenstate state of well-defined angular momentum satisfies (\ref{eq:bohrsommerfeld}) with $I = L$, $\chi = \alpha$ and $j = \ell$, provided also $N - |\ell| \le 2N$ is even.
This is transformed into an integral over ellipses on the unit Poincar\'e sphere: $\alpha = \phi/2$, with $\phi$ the azimuthal angle on the sphere, and the factor of $1/2$ cancels when the range is set to be $0\le \phi \le 2\pi$.
The height on the unit sphere corresponding to $L$ is $Z = L/H = L/(N+1)$; the Bohr-Sommerfeld quantization rule becomes $2\pi \ell/(N + 1) = \int_0^{2\pi} Z\rmd \phi$, with the right hand side being interpreted as the solid angle enclosed between the equator and the circle of height $Z$ on the sphere (as in Figure \ref{fig:families} (a)), which is then fixed at values $-N/(N+1), (-N+2)/(N+1), \ldots, +N/(N+1)$.
This integration path can be visualized as the family of ellipses shown in Figure \ref{fig:families} (c), all with the same size (intensity), shape (angular momentum), but with varying $\alpha$ (major axis direction).
The pattern is circularly symmetric, with two concentric circular caustics enveloping the ellipse family; these correspond semiclassically to LG modes, whose intensity patterns are concentric bright rings with a hole in the middle (unless $\ell = 0$, in which case the ellipse family corresponds to straight lines (swinging orbits) of varying $\ell$).

In order to work mathematically with ellipse families such as this, it is useful to define a function $F$ of $x$ and $y$, parametrized by $H$ and the point on the unit sphere $\bs{u} = (X,Y,Z)$, such that $F = 0$ is the equation for the corresponding ellipse on the Poincar\'e sphere (which is insensitive to the ellipse handedness).
We start with the familiar equation for an ellipse, with semiaxes $a$ and $b$, which satisfies
\begin{equation}
   \frac{x'^2}{a^2} + \frac{y'^2}{b^2} = 1,
   \label{eq:elldef}
\end{equation}
in terms of Cartesian coordinates $x',y'$.
If the major axis of the ellipse is oriented at $\alpha$ to the $x$-axis, $x' = x \cos \alpha + y \sin\alpha$, $y' = y \cos \alpha - x \sin \alpha$.
Also, on the unit Poincar\'e sphere, $2\alpha = \phi = \arctan(Y/X)$, in terms of the sphere's azimuthal angle and components of $\bs{u}$.
Therefore (\ref{eq:elldef}) can be rearranged to give
\begin{equation}
   2 a^2 b^2 = (a^2 + b^2) (x^2 + y^2) + \frac{(a^2-b^2)}{\sqrt{1-Z^2}} \left[ X(x^2-y^2) + Y(2 x y)\right].
   \label{eq:elldef2}
\end{equation}
Now, the ellipse with semiaxes $a$ and $b$, in $x',y'$ components, has $\bs{E} = (a,\rmi b)\exp(\rmi t_0)$ for suitably chosen $t_0$; therefore it has Stokes parameters $S_0 = 2H = a^2+b^2$ and $S_3 = 2 H Z = 2 a b$.
Substituting appropriate expressions for $H$ and $Z$ in place of $a$ and $b$ in (\ref{eq:elldef2}) gives 
\begin{equation}
   F(x,y,H,\bs{u}) = (x^2 + y^2) - X (x^2 - y^2) - Y (2 x y) - H Z^2.
   \label{eq:Fdef}
\end{equation}
For fixed choice of $H$ and $\bs{u}$, $F = 0$ is the equation for an ellipse with axes $Z H^{1/2}/(1\pm \sqrt{1-Z^2})^{1/2}$ and major axis angle $\alpha = \tfrac{1}{2}\arctan(Y/X)$.
By fixing a value of the Hamiltonian $H$, setting $F = 0$ with varying $\bs{u}$ sweeps out families of ellipses as contours in $x,y$. 
For $\bs{u}(h)$ with parameter $h$, the caustic curve satisfies $F = \partial F/\partial h = 0$.
In the case of the family of ellipses with fixed shape and varying $\alpha$, the caustics are concentric circles whose radii given by the ellipse axes, $Z H^{1/2}/(1\pm \sqrt{1-Z^2})^{1/2}$, and the quantization rule fixes the allowed values of $H$ and $Z$. 

A similar argument follows for the quantization of states with definite $M$; the quantization rules are the same, but now the curves are circles around the horizontal axis on the Poincar\'e sphere corresponding to $M$, with fixed $X$ (Figure \ref{fig:families} (b)).
The family of ellipses swept out by this curve is bounded by a rectangle (Figure \ref{fig:families} (e)), and the caustic curves are $x = \pm \sqrt{H}\sqrt{1-X}, y = \pm \sqrt{H} \sqrt{1+X}$.
This is the ellipse family for the HG modes, whose intensity pattern is a grid of horizontal and vertical lines.

The semiclassical picture of the GG modes is the same, as the relevant operator (a linear combination of the constants of motion operators) has the same spectrum as $\widehat{L}$ and $\widehat{M}$: a GG mode corresponds to the circles centred around this direction on the Poincar\'e sphere (Figure \ref{fig:families} (c)).
The caustic envelope of ellipses in this case (Figure \ref{fig:families} (f)), however, is more complicated, and does not correspond to the contours of a coordinate system in which the quantum wavefunction is separable, as indeed the GG beams are not, and must be written in superpositions of the form (\ref{eq:GG}), rather than as separable products of functions in some coordinate system.

This discussion has shown that the connection between the Poincar\'e sphere of classical harmonic oscillator orbits and Gaussian beams is best approached semiclassically, by associating Gaussian beams not with a single point on the sphere, but with a family of classical orbits on a circle whose axis is specified by the point on the sphere, and whose radius is determined by the Bohr-Sommerfeld condition, known to be exact for harmonic oscillators.
The structure of the caustics of the ellipse family determine the structure of the Gaussian modes, and in particular are circular and rectangular for LG and HG beams.

The families of ellipses in Figure \ref{fig:families} (d)-(f), which semiclassically represent LG, HG and GG modes respectively, correspond to all of the classical orbits with this particular value of $H$ and $L, M$ or $X M+Y\overline{M}+Z L$; in the original Hamiltonian phase space, these values are on tori, as discussed in Section \ref{sec:poincare}, and the ellipse families plotted are, in fact, projections of these tori onto 2D configuration space $(Q_x,Q_y)$.
The fact that they are tori is evident: the circularly symmetric case of Figure \ref{fig:families} (d) is a projection down the torus axis, (e) is a side view and (f) is a typical view without any special symmetries.

A full discussion of separability of Gaussian wavepackets is beyond the scope of the present article, and can be approached by a reformulation of the mechanical problem in terms of the Hamilton-Jacobi equation.
It has been proved \cite{boyer:1975Lie6,boyer:1975Lie7} that there are three coordinate systems in which the quantum 2D isotropic harmonic oscillator is separable: plane polar coordinates (giving the LG modes), Cartesian coordinates (giving the HG modes), and elliptic coordinates, giving the Ince-Gaussian modes \cite{Bandres:2004Ince,Schwarz:2004observation} (the latter is a family of coordinate systems, parametrized by the inter-focal distance $f$, smoothly interpolating between polar and Cartesian coordinates), which are products of Ince polynomials \cite{arscott:periodic,boyer:1975Lie7} in elliptic coordinate variables.
The underlying classical dynamics of the Ince-Gauss modes will be explored elsewhere; the corresponding classical constant of the motion is $L^2 + g M$ for real parameter $g$, which is not a linear combination of the $\rm{su}(2)$ basis of the Poisson algebra, like the GG modes.

\section{Concluding remarks}

We have explored the connection between the Poincar\'e spheres of polarization (classical isotropic 2D harmonic oscillators) and Gaussian laser beams; these connections were implicit in the work of Danakas and Aravind \cite{Danakas:1992analogies}, which also celebrates its 25th anniversary this year.
We have also presented the viewpoint that the Poincar\'e sphere picture emerges from the `swings and roundabouts' picture of the 2D oscillator, with linear orbits with momenta of unspecified sign, but circular orbits of definite sign, as emphasized for LG beams by \cite{Allen:1992orbital}.
By examination of the semiclassical connection between the classical and quantum pictures, we have seen that the `Poincar\'e sphere' of Gaussian modes can be understood by families of ellipses on circles around this axis, with the circles fixed by a semiclassical quantization rule.

The familiarity to most contemporary physicists of the quantum mechanical formalism has allowed us to present these connections without explanation of the true underlying optical physics -- mathematical procedures from quantum mechanics such as canonical quantization can be followed without physical understanding.
One might be tempted to claim, especially in the light of the semiclassical picture, that the ellipses in question are `rays', but how should elliptical rays be interpreted?

A natural explanation (which will not be further formalized here) is that a Gaussian beam in three dimensions are made up of a 2-parameter families of rays, crossing the focal plane at a definite position $\bs{Q}$ and with a definite inclination $\bs{P}$.
The ellipses $\bs{Q}(t)$ of (\ref{eq:Qdef}) represent one-parameter families of these rays, lying on a hyperboloid with transverse elliptic cross-section: this hyperboloid is clearly self-similar on propagation.
The one-parameter families of ellipses in Section \ref{sec:semic} provides the extra parameter determining the families of rays building up the Gaussian beam, so the caustics are the envelopes of the classical rays of the Gaussian beams in this picture.
Semiclassically quantizing different constants of classical motion guarantees the ray families have definite properties (e.g.~angular momentum, definite Hamiltonian (mode order) in $x$ and $y$, etc); each elliptic 1-parameter ray family is given a phase, and the quantization condition required this phase must be singlevalued around the ellipse family.
Moving semiclassically from the ray picture to the wave picture involves decorating the caustics with appropriate diffraction catastrophes \cite{nye:natural}---primarily Airy functions (and Pearcey functions in the case of Figure \ref{fig:families} (c))---and approximations involving diffraction catastrophe integrals improve as $N $ becomes asymptotically large.
Viewing the ellipses as contours of $F$ implies the caustic construction has difficulty distinguishing the sign of angular momentum, but has no such difficulty for linear orbits -- another distinction between swings and roundabouts.
We will describe this construction in more detail elsewhere \cite{ad:gaussian}.

Even ignoring these details, the analysis here shows the utility of the operator-based, `Heisenberg picture' approach to optical beams \cite{Stoler:1981operator}, where the wave forms of structured light are identified as eigenfunctions of certain operators, which commute with an overall `Hamiltonian' operator such as the harmonic oscillator Hamiltonian for Gaussian beams, or minus the transverse Laplacian $-\nabla_{\bot}^2 = - \partial_x^2 - \partial_y^2$ for propagation-invariant beams.
In this latter case, we see Bessel beams as eigenfunctions of the angular momentum operator $\widehat{L}$, and so on.
In the limit $N \to \infty$, Gaussian beams approach diffraction-free beams, so LG beams become Bessel beams in the limit $p \to \infty$.
For propagation-invariant beams, the counterpart of $\widehat{M}$ is simply the difference of momenta $\widehat{p}_x^2 - \widehat{p}_y^2$ (i.e.~there is no potential), and the eigenfunctions are standing wave patterns such as $\cos(k_x x)\sin(k_y y)$, which is the appropriate limit of HG beams.
In this picture, Mathieu beams \cite{GutierrezVega:2000alternative}, the propagation-invariant beams separable in elliptic coordinates, are eigenfunctions of $\widehat{L}^2 + g (\widehat{p}_x^2 - \widehat{p}_y^2)$ (for constant $g$) are the appropriate high-$N$ limit of Ince-Gaussian beams.
Diffraction-free beams which are eigenfunctions of $\widehat{L} + g (\widehat{p}_x^2 - \widehat{p}_y^2)$, counterparts to GG beams, do not seem to have been  discussed significantly in the literature; although there are various Poincar\'e sphere-like constructions, only for the Gaussian beams does the semiclassical structure and connection between classical and wave mechanical pictures extend so deeply.

In terms of operators, therefore, the HG, LG and GG beams are interrelated as eigenfunctions of a family of operators parametrized by points on the Poincar\'e sphere, and hence are related to each other by abstract rotations (\rm{SU}(2) transformations) in 3D Stokes space (the space of constants of motion, linear combinations of $L, M$ and $\overline{M}$).
Despite the close mathematical links with rotation in real space, the only spatial rotations admitted here are 2D, $\rm{SO}(2)$ rotations about the axis, i.e.~rotations of $\alpha$; the other $\rm{SU}(2)$ transformations, changing the shape of the ellipse (corresponding, in polarization, to the action of a phase plate) are equivalent to transformations of a `hidden symmetry' (ultimately, to the superintegrability of the 2D oscillator), meaning that the problem is separable in multiple coordinate systems.
The 2D oscillator is an important example, as is the 3D hydrogen atom with a hidden $\rm{SO}(4)$ symmetry \cite{Fradkin:1967existence}, the extra classical constant of the motion being the Runge-Lenz vector (which Poisson commutes with the angular momentum vector), leading to the degeneracy of the quantum hydrogen atom, and its separability in both spherical polar coordinates and parabolic cylindrical coordinates \cite{merzbacher:quantum}.

Unlike a harmonic potential, the diffraction-free Hamiltonian $-\nabla^2_{\bot}$ (with eigenvalues $k^2_{\bot}$) commutes with more operators than the harmonic oscillator Hamiltonian,  such as the linear momentum operator in some specific direction $-\rmi(k_x \partial_x + k_y \partial_y$), whose eigenfunctions are travelling plane waves $\exp(\rmi k_x x + \rmi k_y y)$, distinct from standing plane waves which are eigenfunctions of the counterparts of $\widehat{M}$, $\widehat{\overline{M}}$.
Families of propagation-invariant beams can be constructed as eigenfunctions of interpolations of linear momentum and angular momentum, such as the so-called pendulum beams \cite{Dennis:2013propagation}, which are eigenfunctions of $\widehat{L}^2 - g \widehat{p}_y$, waves whose motion is a hybrid of standing circular and travelling linear motion, and have the special physical property of minimizing a certain natural angular momentum uncertainty relation \cite{Forbes:2003uncertainty}.
Many beams in the menagerie of structured light modes can be formulated in this way, and it is likely that there will be many other new and interesting forms of structured light which emerge naturally from physically-interesting operators.\vskip6pt

\acknowledgments{The authors are grateful for discussions over several years with many colleagues on this subject, particularly Eugene Abramochkin, Andrea Aiello, Les Allen, Stephen Barnett, Michael Berry, Konstantin Bliokh, Johannes Courtial, Oliver Dyer, Greg Forbes, J\"org G\"otte, Julio Guti\'errez-Vega, Stephen Habraken, John Hannay, Peter Jones, Jari Lindberg, Michael Morgan, Bill Miller, Gerard Nienhuis, Miles Padgett, Ulrich Schwarz, Danica Sugic and Bernardo Wolf.
The Poincar\'e translation was assisted by Caroline Turner.
MRD acknowledges support from the Royal Society and the Leverhulme Trust during this work. 
MAA acknowledges support from the National Science Foundation (PHY 1507278).
}

\end{document}